\documentclass[twocolumn]{revtex4}
\usepackage{epsfig,bm,dcolumn}
\usepackage{graphicx}
\usepackage{epsfig}
\usepackage{latexsym}
\usepackage{bm}
\usepackage{amsmath}
\usepackage{lipsum}

\begin{document}

\renewcommand{\ni}{{\noindent}}
\newcommand{\dprime}{{\prime\prime}}
\newcommand{\be}{\begin{equation}}
\newcommand{\ee}{\end{equation}}
\newcommand{\bea}{\begin{eqnarray}} 
\newcommand{\eea}{\end{eqnarray}}
\newcommand{\nn}{\nonumber} 
\newcommand{\bk}{{\bf k}}
\newcommand{\bQ}{{\bf Q}}
\newcommand{\q}{{\bf q}}
\newcommand{\s}{{\bf s}}
\newcommand{\bN}{{\bf \nabla}}
\newcommand{\bA}{{\bf A}}
\newcommand{\bE}{{\bf E}}
\newcommand{\bj}{{\bf j}}
\newcommand{\bJ}{{\bf J}}
\newcommand{\bs}{{\bf v}_s}
\newcommand{\bn}{{\bf v}_n}
\newcommand{\bv}{{\bf v}} 
\newcommand{\la}{\langle}
\newcommand{\ra}{\rangle} 
\newcommand{\dg}{\dagger}
\newcommand{\br}{{\bf{r}}} 
\newcommand{\brp}{{\bf{r}^\prime}} 
\newcommand{\bq}{{\bf{q}}}
\newcommand{\hx}{\hat{\bf x}} 
\newcommand{\hy}{\hat{\bf y}}
\newcommand{\bS}{{\bf S}} 
\newcommand{\cU}{{\cal U}}
\newcommand{\cD}{{\cal D}} 
\newcommand{\bR}{{\bf R}}
\newcommand{\pll}{\parallel} 
\newcommand{\sumr}{\sum_{\vr}} 
\newcommand{\cP}{{\cal P}} 
\newcommand{\cQ}{{\cal Q}} 
\newcommand{\cS}{{\cal S}}
\newcommand{\ua}{\uparrow} 
\newcommand{\da}{\downarrow}

\title{Unconventional superconductivity in a strongly correlated band-insulator without doping}

\vspace{-1.0cm}
\author{Anwesha Chattopadhyay$^{1}$, H. R. Krishnamurthy$^{2}$, Arti Garg$^{1}$ } 
\affiliation{$^{1}$ Condensed Matter Physics Division, Saha Institute of Nuclear Physics, HBNI, 1/AF Bidhannagar, Kolkata 700 064, India\\
  $^{2}$ Centre for Condensed Matter Theory, Department of Physics, Indian Institute of Science, Bangalore 560 012, India}

\begin{abstract}
  We present a novel route for attaining unconventional superconductivity (SC) in a strongly correlated system {\it without doping}. In a simple model of a {\it correlated band insulator} (BI) at {\it half-filling} we demonstrate, based on a generalization of the projected wavefunctions method, that SC emerges when e-e interactions and the bare band-gap are both much larger than the kinetic energy, provided the system has sufficient frustration against the magnetic order. As the interactions are tuned, SC appears sandwiched between the correlated BI followed by a paramagnetic metal on one side, and a ferrimagnetic metal, antiferromagnetic (AF) half-metal, and AF Mott insulator phases on the other side.
  \end{abstract}

\maketitle

The discovery of unconventional superconductivity in a variety of materials, such as high $T_c$ superconductivity in cuprates~\cite{Bednorz}, iron pnictides and chalcogenides~\cite{pnictide_expt}, in organic superconductors~\cite{organic}, heavy fermions~\cite{HF} and very recently in magic angle twisted bilayer graphene~\cite{MTBLG,MTBLG2}, has always ignited worldwide interest owing to their rich phenomenonology, the theoretical challenges they pose, scientific implications and broad application potential. In almost all of these examples, superconductivity appears upon {\it chemically doping} a parent compound  {\it away from commensurate filling}~\cite{Bednorz,pnictide_expt,Lee,Pnictides,MTBLG,MTBLG2}, though in some cases inducing charge fluctuations by changing pressure also leads to the superconducting phase~\cite{organic,Pnictides}. An important experimental fact is that chemical  doping inevitably induces disorder, as is clearly the case in high $T_c$ superconductors, which makes these materials very inhomogeneous~\cite{Pan,Mcelroy,Garg,Tang}. It is a theoretical and experimental challenge to come up with new mechanisms and materials for {\it clean} high $T_c$ superconductors.

Strong e-e correlations are crucial for unconventional superconductivity (SC). In most of the known unconventional superconductors~\cite{Bednorz,pnictide_expt,organic,MTBLG,MTBLG2,Lee,Pnictides} the low temperature phase of the parent compound is either a strongly correlated antiferromagnetic (AF) Mott insulator where charge dynamics is completely frozen, or a AF spin-density-wave phase with at least moderately strong correlations.
But the possibility of a SC phase in a {\it strongly correlated band-insulator} has been explored very little so far, either theoretically or experimentally.

In this work, we show how an AF spin-exchange mediated SC can be realized {\it without doping} in a simple model of a strongly correlated band insulator (BI), where the bare band 
   \begin{figure}[h!]
\begin{center}
\vspace*{0.3cm}
\hspace*{-0.75cm}
\includegraphics[scale=0.33,angle=-90]{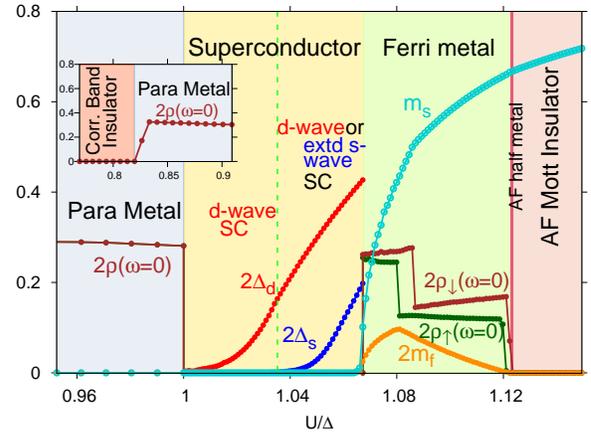}
\caption{Zero temperature phase diagram for the ionic Hubbard model (see text) on a square lattice with e-e interaction $U=10t$ and $t'=0.4t$ as a function of the  interaction to bare band-gap ratio ($U / \Delta$). For $ \Delta \gg U \gg t$, the system is a correlated BI without any magnetic order. On increasing $U/\Delta$, first the gap in the single particle excitation spectrum closes, as shown by the non-zero single particle density of states at the Fermi energy $\rho(\omega=0)$, resulting in a metallic phase. On further increasing $U/\Delta$, SC sets in and lasts over a significanly broad range ($\Delta \in [9.3:10]t$) before ferrimagnetic order with non-zero staggered ($m_s$) and uniform ($m_f$) magnetization sets in via a first order transition. This is a Ferri {\it metal} phase with non-zero, spin asymmetric, spectral density at the Fermi-level: $\rho_\uparrow(\omega=0) \neq \rho_\downarrow(\omega=0) >0$. As $U/\Delta$ increases further, $m_f \rightarrow 0$ whence the magnetic order becomes AF and an up-spin spectral gap opens up such that $\rho_\uparrow(\omega=0)=0$  while the down-spin electrons  are still conducting, resulting in a sliver of AF {\it half-metal}. Eventually the system becomes an AF Mott insulator as $U/\Delta$ increases further.}
\vskip-1cm
	\label{pd1}
\end{center}
    \end{figure}
   gap ($\Delta$) and the e-e interaction ($U$) both dominate over the kinetic energy. As $U$ is increased (but still remains of the order of $\Delta$), the single particle excitation gap in the BI closes, resulting in a  metallic phase. Upon further increasing $U$, SC develops by the formation of a coherent macroscopic quantum condensation of electron pairs, provided the metal has enough low energy quasiparticles and the system has enough frustration against the magnetic order.  The SC features tightly bound short coherence length Cooper pairs with a $T_c$ well separated from the energy scale at which the pairing amplitude builds up. The phase diagram, whose section with all model parameters fixed except for $U/\Delta$ is shown in Fig.1, presents a plethora of exoctic phases in the vicinity of a broad region of the SC phase. 

Our starting point is a variant of the Hubbard model, known as the {\it ionic} Hubbard model (IHM), where, on a bipartite lattice with sub-lattices A and B, a staggered
ionic potential $\Delta/2$ is present in addition to coulomb repulsion ($U$):
\bea
\mathcal{H}=&-\sum_{i,j\sigma}(t_{ij}c_{i\sigma}^{\dagger}c_{j\sigma} + h.c.)-\mu\sum_{i}n_{i} \nonumber\\
    &-\frac{\Delta}{2}\sum_{i \in A} n_i +\frac{\Delta}{2}\sum_{i \in B} n_i +U\sum_{i}n_{i\uparrow}n_{i\downarrow}
    \label{model}
    \eea
The amplitude for electrons with spin $\sigma$ to hop between sites $i$ and $j$ is $t_{ij}=t$ for near-neighbours and $t_{ij}=t^\prime$ for second neighbours. The chemical potential $\mu$ is chosen to fix the average occupancy at $n = 1$ (half-filling). The staggered potential doubles the unit cell, and (for $t' < \Delta/4$) induces a gap between the two electronic bands that result, making the system a BI for $U=0$.

    The parameter range  of interest for this work is $U\sim\Delta\gg t,t'$, where a theoretical solution can be obtained based on a generalization of the projected wavefunctions method~\cite{zhang,Arun,edegger-advphysics,vanilla,Ogata,Lee2,Anwesha1}. In this limit, at half-filling holons (doublons) are energetically expensive on the $A$ ($B$) sites and can be projected out of the low energy Hilbert space. Consequently, though all hopping processes connecting the low and high energy sectors of the Hilbert space are eliminated, the system still has charge dynamics through first neighbor hopping processes such as $|d_A h_B\ra \Leftrightarrow |\uparrow_A\downarrow_B\ra$ (with $d$ representing a doublon and $h$ a holon) and second neighbour hopping processes which allow doublons (holons) to hop on the A (B) sublattice~\cite{hubbard}.
    
    The effective low energy Hamiltonian at half-filling, $H_{eff}$, is an extended $t-t^\prime-J-J^\prime$ model acting on a projected Hibert space:
    \vskip-0.5cm
    \[H_{eff}=\mathcal{P}\bigg[\bigg(-t\sum\limits_{<ij>,\sigma}c^\dagger_{iA\sigma}c_{jB\sigma}-t^\prime\sum\limits_{\substack{<<ij>> \\ \alpha,\sigma}}
    c^\dagger_{i\alpha\sigma}c_{j\alpha\sigma}+h.c.\bigg)\]
    \vskip-0.83cm
    \[+J^\prime\sum\limits_{<<ij>>}\sum\limits_{\alpha}S_{i\alpha}.S_{j\alpha}-\frac{1}{4}(2-n_{iA})(2-n_{jA})-\frac{1}{4}n_{iB}n_{jB}\] \vskip-0.83cm 
    \[+J\sum\limits_{<ij>}\bigg(S_{iA}.S_{jB}-(2-n_{iA})n_{jB}/4\bigg)-\mu\sum_in_i \] \vskip-0.83cm
    \be
     +H_0+H_d+H_{tr}+...\bigg]\mathcal{P} 
    \label{tj}
    \ee
    Here $J= 2t^2/(U+\Delta)$ and $J^\prime=4{t^\prime}^2/U$.
    $H_0$ is the rescaled Hubbard interaction term in the projected Hilbert space and $H_d$ ($H_{tr}$) indicates other dimer (trimer) processes.
    We treat the projection constraint in $H_{eff}$ using the generalised Gutzwiller approximation~\cite{Anwesha1} and solve it using a renormalized Bogoliubov mean field theory. Gutzwiller approximations~\cite{edegger-advphysics,Ogata,Anwesha1} of the sort we use  have been well vetted against quantum Monte Carlo calculations~\cite{Arun,vanilla} and dynamical mean field theory~\cite{Anwesha1}. Details of this {\it renormalized mean field theory}, Gutzwiller approximation and the various terms in $H_{eff}$ are given in the Supplementary Material (SM)~\cite{SM}. 

Our main findings are summarised in the phase diagram of Fig.~1, which shows a linear section (along the $U / \Delta$ axis) of the full phase diagram in Fig. 2[e], for the IHM on a 2d square lattice at $t^\prime=0.4t$.
The correlated BI, stable for $\Delta \gg U \gg t$, is paramagnetic and adiabatically connected to the BI phase of the non-interacting IHM. As $\Delta$ approaches $U$, the low energy hopping processes ($|d_A h_B\ra \Leftrightarrow |\uparrow_A\downarrow_B\ra$) become more prominent, increasing charge-fluctuations such that the gap in the single particle excitation spectrum closes, leading to a paramagnetic metallic (PM) phase with finite single particle density of states (DOS) $\rho(\omega=0)$ at the Fermi energy, though for most of the parameter regime the PM phase is a compensated semi-metal with small Fermi pockets (Fig. 3 and following discussion)~\cite{PM}.
On further increasing $U/\Delta$, in the presence of sufficiently large $t^\prime$, robust SC sets in for $\Delta \sim U$ over a broad range of $U/t$ (as shown in SM) due to the formation of coherent Cooper pairs of quasi-particles which live near the Fermi pockets, and survives for a sizeable range of $\Delta$ ($\in [9.3:10]t$). The pairing amplitudes $\Delta_{d/s}$ for both the pairing symmetries we have studied, namely,  the d-wave and the extended s-wave, increase monotonically with $U / \Delta$ and drop to zero via a first order transition at the transition to the ferrimagnetic metal~\cite{coexist}.

The  ferrimagnetic metal (FM) phase is characterised by non-zero values of the staggered and uniform magnetizations $m_{s,f}=(m_A\mp m_B)/2$ with $m_{A/B}$ being the sublattice magnetizations, along with finite spin asymmetric DOS $\rho_\sigma(\omega=0)$ at the Fermi energy. With further increase in $U/\Delta$  the FM evolves into an AF {\it half-metal} phase in which the system has only staggered magnetization (i.e., $m_f=0$) and the single particle excitation spectrum for up-spin electrons is gapped while the down-spin electrons are still in a semi-metal phase. Eventually, for a large enough $U/\Delta$, both spin spectra become gapped - the system becomes an AF Mott insulator~\cite{note}. 

We next discuss the changes in the behavior of the  system with increasing $U/\Delta$ for varying values of $t^\prime$, as depicted in Fig. 2. For $t^\prime=0$, the system shows a direct first order transition from an AF ordered phase to a correlated BI with a sliver of a half-metallic AF phase close to the AF transition point, consistent with most other theoretical work in this limit~\cite{Anwesha2,watanabe,Soumen,Dagotto} barring one exception~\cite{Rajdeep}. When $t^\prime$ is non-zero but small, due to the breaking of particle-hole symmetry as well as the frustration induced by the second neighbour spin-exchange coupling $J^\prime$, the system first attains ferrimagnetic order for a range of $U/\Delta$, beyond which it has pure AF order as shown in panels (a,b) of Fig. 2. The magnetic transition occurs at increasingly larger values of $U/\Delta$ with increasing $t^\prime$ (except for an initial decrease for small values of $t^\prime$), which helps in the development of a stable SC phase.

To stabilize the SC phase, a minimum threshold value of $t^\prime$ (which is a function of $U$) is required, partly in order to frustrate the magnetic order as mentioned
above, but more importantly to gain sufficient kinetic energy by intra-sublattice hopping of holons and doublons on their respective sublattices where they are energetically allowed. 
\begin{figure}[h!]
  \begin{center}
    \includegraphics[scale=0.33,angle=-90]{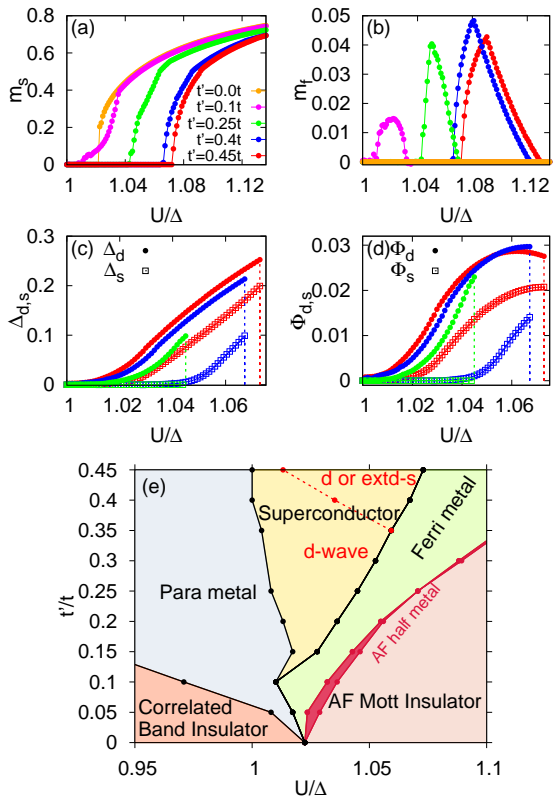}
    \caption{Top panels show the staggered and uniform magnetization as functions of $U / \Delta$ for several values of $t'$ and $U=10t$. With increasing $t'$, the magnetic transition point first decreases for $t^\prime \le 0.12t$, and then starts increasing again. The magnetic transition is of first order for $t^\prime=0$ as well as for large values of $t^\prime$, though for intermediate values of $t^\prime$ the magnetization tuns on continuously. Panel (c) shows the SC pairing amplitude $\Delta_{d/s}$, for the d-wave and extended s-wave pairing symmetry. With increasing  $t^\prime$ the range in $U/\Delta$ over which the SC is stable gets wider, and the amplitudes of both d-wave and extended s-wave pairings get enhanced. Panel (d) shows the SC order parameter $\Phi_{d/s}$, which also gives an estimate of the SC transition temperature, $T_{c}$.
      The bottom panel (e) shows the complete zero temperature phase diagram for $U=10t$ in the $t^\prime/t$-$U/\Delta$ plane.}
    \vskip-1cm
\label{pd2}
\end{center}
\end{figure}
While a stable d-wave SC phase turns on for $t^\prime > 0.1t$ for $U=10t$, as shown in Fig. 2, SC in the extended s-wave channel gets stabilized for much larger value of $t^\prime > 0.35t$ . In an intermediate regime of $U/\Delta$ and $t^\prime$, states with both d-wave and extended s-wave symmetry are viable solutions with energies that are very close (See SM for details). As $t^\prime$ increases, the pairing amplitude increases and the range of $U/\Delta$ over which the SC phase exists becomes broader for both the pairing symmetries studied~\cite{note2}.

The SC order parameter $\Phi_{d/s}$ is defined in terms of the  off-diagonal long-range order in the correlation function $F_{\gamma_1\gamma_2}({\bf{r}}_i-{\bf{r}}_j) =\la B^\dagger_{i\gamma_1}B_{j\gamma_2}\ra \rightarrow \Phi_{\gamma_1} \Phi_{\gamma_2}$ as $|{\bf{r}}_i-{\bf{r}}_j| \rightarrow \infty$, where $B^\dagger_{i\gamma}$ creates a singlet on the bond $(i,i+\gamma)$. Fig. 2 shows the SC order parameter, which has been obtained after taking care of renormalization required in the projected wavefunction scheme (see SM).
Since the SC order parameter for this system is much smaller than the strength of the pairing amplitude, with increase in temperature the SC will be destroyed at $T_c$ by the loss of coherence among the Cooper pairs, leaving behind a pseudo-gap phase with a soft gap in the single particle density of states due to the Cooper pairs which will exist even for $T > T_c$. Thus $\Phi_{d/s}$ also provides an estimate of the SC transition temperature $T_c$. The maximum estimated $T_{c}$ for $U=10t$ on a square lattice is approximately $0.03t$ for the d-wave SC phase, which for a hopping amplitude comparable to that in cuprates ($t\sim 0.4 eV$) gives a $T_c\sim 150K$, and there is a considerable scope for enhancing $T_c$ by tuning $U/\Delta$ as well as $t^\prime$.
\begin{figure}[h!]
  \begin{center}
    \hspace{0cm}
    \vskip-0.4cm
    \includegraphics[scale=0.13]{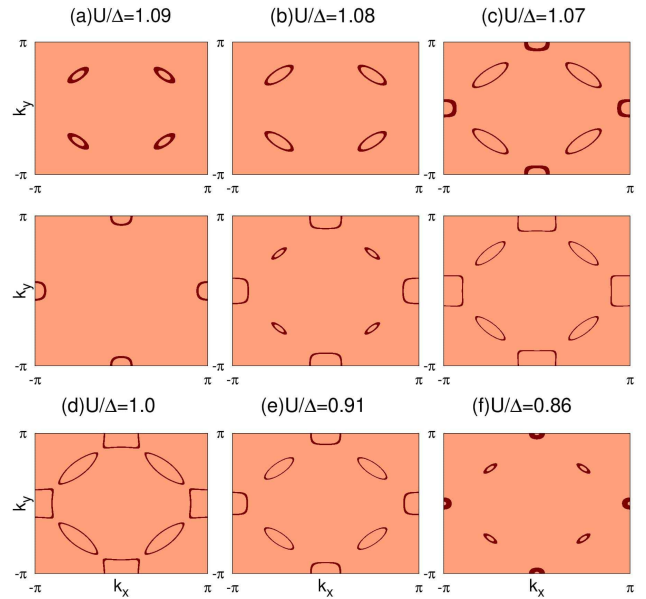}
    \vskip-0.5cm
\caption
    {The top two rows show the spin resolved low energy spectral functions $A_\sigma(k,\omega \sim 0)$ (integrated over $|\omega| \le (0.01-0.02)t$ for a $3000 \times 3000$ system) in the full BZ in the FM phase for $t'=0.35t,U=10t$, with up-spin (down-spin) spectral functions shown in the first (second) row. At $U/\Delta=1.09$, the up spin channel has electron pockets while the down spin channel has small hole pockets. As $U/\Delta$ decreases, these Fermi pockets become bigger, the down (up) spin spectral function gets additional electron (hole) pockets. The last row shows $A(k,\omega \sim 0)$ (same for up or down spins) for the PM phase. Moving towards the SC phase by increasing $U/\Delta$, Fermi pockets in the PM state go on expanding until they almost start touching each other, at which point the SC sets in by formation of Cooper pairs between electrons close to the Fermi energy.}
\label{Fig3}
\end{center}
\vskip-0.4cm
\end{figure}

A striking feature of the phase diagram in Fig. 2 is that, though the origin of SC in this model is due to the AF spin-exchange interactions, SC sets in only after the system has evolved to a PM or a FM phase. In order to understand the charge dynamics as the system approaches the SC phase, we have analysed the single particle spectral functions $A_\sigma(k,w\sim 0)$ which can be directly measured in angle resolved photoemission spectroscopy (ARPES). Fig.~3 shows $A_{\sigma}(k,\omega \sim 0)$, non-zero values of which determine the energy contours on which low energy quasiparticles live in the Brillouin zone (BZ) (see SM for details).
Panels (a-c) show $A_{\sigma}(k,w\sim 0)$ in the FM phase for which the up-spin channel has electron pockets around the points ${\bf{K}}=(\pm \pi/2,\pm \pi/2)$ in the BZ and the down spin channel has small hole pockets around the points ${\bf{K}}^\prime = (\pm \pi,0),(0,\pm \pi)$ (see SM for details), as shown in panel (a). As $U/\Delta$ decreases within the FM phase, and  approaches the SC phase, the electron pockets (hole-pockets) in the up-spin (down-spin) spectral function become bigger, and the down-spin channel gets additional electron pockets while the up-spin channel gets additional hole pockets, as shown in panel (c).

In the PM phase, $A(k,w\sim 0)$ has spin symmetric electron pockets (around ${\bf{K}}$) and hole pockets (around ${\bf{K}}^\prime$). As $U/\Delta$ increases through the PM phase, these Fermi pockets slowly expand such that they almost touch each other before the system enters into the SC phase. Similar behaviour is seen with an increase of $t^\prime$ in the PM or the FM phases (see SM for details).
\begin{figure}[h!]
  \begin{center}
    \vskip-0.4cm
   \includegraphics[scale=0.35,angle=-90]{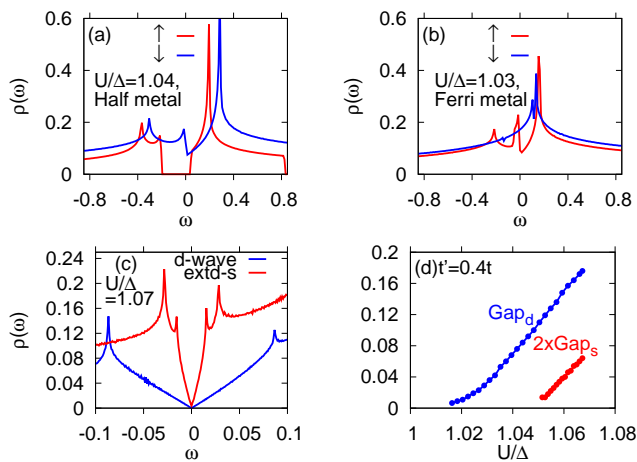}
  \caption{Panels (a)-(b) show the spin resolved single particle density of states (DOS) $\rho_\sigma(\omega)$ for $t'=0.15t$ and $U=10t$. At $U/\Delta\sim 1.04$, $\rho_\downarrow(\omega=0)$ is finite where as $\rho_\uparrow(\omega=0)=0$ with a finite spectral gap, corresponding to the AF half-metal phase. At $U/\Delta=1.03$, the DOS at the Fermi energy is finite in both the spin channels but $\rho_{\uparrow}(\omega) \neq \rho_{\downarrow}(\omega)$ corresponding to the FM phase.
    Panel (c) shows $\rho(\omega)$ for the  d-wave and extd-swave SC phases for $U=10t$ and $t'=0.4t$. $\rho(\omega)$ shows a linear increase with $|\omega|$ for $\omega \sim 0$ for both the SC phases. Panel (d) shows the gap in the DOS, which is the peak to peak distance in $\rho_\sigma(\omega)$, for both the d-wave and the extended s-wave pairing symmetries. }
  \vskip-1cm
\end{center}
\end{figure}
\vskip-0.4cm
Fig.~4. shows the spin-resolved DOS $\rho_\sigma(\omega)$ vs $\omega$ which provides additional evidence for the existence of various metallic phases as depicted in the phase diagram in Fig.~2. The para metal, ferri-metal and the AF half-metal phases are all compensated semi metals, which is reflected in the depletion in the DOS at the Fermi energy and is  consistent with the small Fermi pockets shown in Fig.~3.
We have also analysed the DOS in the SC phase. As shown in Fig. 4[c], $\rho(\omega\sim 0)\sim |\omega|$ which is a signature of the gapless nodal excitations in the d-wave SC phase. Interestingly, even for the extended s-wave SC phase $\rho(\omega\sim 0) \sim |\omega|$ as the pairing takes place around the small Fermi pockets centered at the ${\bf{K}}$ or ${\bf{K}}^\prime$ points in the BZ, where the pairing amplitude $\Delta_s(k)=\Delta_s(\cos(k_x)+\cos(k_y))$ has nodes as well, resulting in gapless excitations.  
The gap, which is the peak to peak distance in the DOS, is much larger in the d-wave SC phase than in the extended s-wave phase, consistent with the former being the stable phase . Infact for the extended s-wave phase, $Gap_s$ is  only slightly larger than the SC order parameter $\Phi_s$, which indicates that the extended s-wave SC phase will have a narrower pseudogap phase above $T_c$, compared to the d-wave case.     

The origin of unconventional SC in most of the materials known today ~\cite{organic,Lee,Pnictides,MTBLG} can be understood in terms of the strongly correlated limit of the paradigmatic Hubbard model (single or multi band) but only upon doping the system away from half-filling~\cite{vanilla,Lee,pwa,kotliar,Pnictides,HRK}.
In the theoretical model we have studied here, SC appears even at half-filling, and therefore without the disorder that inevitably accompanies doping.  A remarkable feature is that the SC phase in this model of a correlated BI is sandwiched between paramagnetic metallic and ferrimagnetic metallic phases, which makes the zero temperature phase diagram very different from that of the known unconventional SCs like high $T_c$ cuprates ~\cite{Lee} or the more recent magic angle twisted bilayer graphene~\cite{MTBLG}. We expect that the SC phase in this model has transition temperatures comparable to those of cuprates and that it also has a pseudogap phase like in cuprates.

The IHM has been realized for ultracold fermions on an optical honeycomb lattice~\cite{IHMexpt}, where the state-of-the art engineering allows the parameters in the Hamiltonian to be tuned with great control. Hence it will be interesting and perhaps the easiest to explore our theoretical proposal in these systems. In the context of the recent developments in layered materials and heterostructures, it is possible to think of many scenarios where the IHM can be used as a minimal model, for example, graphene on h-BN substrate or bilayer graphene in the presence of a transverse electric field~\cite{BLG} which generates the staggered potential. The limit of strong correlation, crucial for realizing the SC phase, can be achieved in these materials by applying a strain or twist. Band insulating systems with two inequivalent strongly correlated atoms per unit cell, frustration in hopping and antiferromagnetic exchange, and lack of particle-hole symmetry, are likely tantalizing candidate materials as well. Our work suggests that the search for such novel materials where superconductivity can be realized at half filling with sufficiently high transition temperatures can perhaps emerge as an exciting, though challenging new research frontier in condensed matter physics.

A. G. and H. R. K gratefully thank the Science and Engineering Research Board of the Department of Science and Technology, India for financial support under grants No. CRG/2018/003269 and SB/DF/005/2017 respectively. A. C. acknowledges financial support by Department of Atomic Energy, Government of India.

\onecolumngrid

\section*{{\Large \bf Supplementary Material}}

\twocolumngrid

\noindent
    {\bf SM A. Large $U$ and large $\Delta$ limit of the ionic Hubbard model}:
    \\
    \noindent
    We solve the model in Eq. (1) of the main paper in the limit $U\sim\Delta\gg t,t'$. In this limit and at half-filling, holons are energetically expensive on the $A$ sites (with onsite potential $-\frac{\Delta}{2}$) and doublons are expensive on the $B$ sites (with onsite potential $\frac{\Delta}{2}$); i.e., in the low energy subspace $h_A$ and $d_B$ are constrained to be zero.  We do a generalized similarity transformation on this Hamiltonian, $\tilde{H}=e^{-iS}He^{iS}$, such that all first and second neighbour hopping processes connecting the low energy sector to the high energy sector of the Hilbert space are eliminated. The similarity operator of this transformation is\small{ $S=-\frac{i}{U+\Delta}({H_{t}^{+}}_{A\rightarrow B}-{H_{t}^{-}}_{B\rightarrow A})-\frac{i}{\Delta}({H_{t}^{0}}_{A\rightarrow B} -{H_{t}^{0}}_{B\rightarrow A})-\frac{i}{U}({H_{t'}^{+}}_{A\rightarrow A}-{H_{t'}^{-}}_{A\rightarrow A})-\frac{i}{U}({H_{t'}^{+}}_{B\rightarrow B}-{H_{t'}^{-}}_{B\rightarrow B}) $ }\normalsize where $H_{t / t'}^{+}$ represents first or second neighbour hopping processes which involve an increase in $h_A$ or $d_B$ by one and  $H_{t / t'}^{-}$ on the other hand represent hopping processes which involve a decrease in $h_A$ or $d_B$ by one. $H_{t}^{0}$ processes do not involve a change in $h_A$ and $d_B$.
 The low energy effective Hamiltonian $H_{eff}$ obtained by this transformation is given in Eq. (2) of the paper, with $H_0= \frac{U-\Delta}{2}\sum_{i}[n_{iA\uparrow}n_{iA\downarrow}+(1-n_{iB\uparrow})(1-n_{iB\downarrow})]$. Further details can be found in ~\cite{Anwesha1}. 
$H_{eff}$ acts on a projected Hilbert space which consists of states $|\Phi\ra = \mathcal{P}|\Phi_0\ra$ where the projection operator $\mathcal{P}$ eliminates components with $h_A \geq 1$ or $d_B \geq 1$ from $|\Phi_0\ra$. We use here the Gutzwiller approximation~\cite{edegger-advphysics,vanilla,Anwesha1} to handle the projection, by writing the expectation value of an operator $Q$ in a state $\mathcal{P}|\Phi_0\ra$ as the product of a Gutzwiller factor $g_Q$ times the expectation value in $|\Phi_0\ra$ so that $\la Q\ra \simeq g_Q\la Q\ra_0$. The standard procedure~\cite{edegger-advphysics} for calculating $g_Q$ has been generalised by us for the case where holons are projected out from one sublattice and doublons from the other ~\cite{Anwesha1}. 

 We thus obtain the renormalized effective Hamiltonian with the inter-sublattice kinetic energy  $\langle c^{\dagger}_{iA\sigma}c_{jB\sigma}\rangle \approx g_{t\sigma} \langle c^{\dagger}_{iA\sigma}c_{jB\sigma}\rangle_{0}$, and intra-sublattice kinetic energy  $\langle c^{\dagger}_{i\alpha\sigma}c_{j\alpha\sigma}\rangle \approx g_{\alpha\sigma} \langle c^{\dagger}_{i\alpha\sigma}c_{j\alpha\sigma}\rangle_{0}$. The inter-sublattice spin correlation $\langle {\bf S}_{iA}\cdot{\bf S}_{jB}\rangle \approx g_{sAB} \langle {\bf S}_{iA}\cdot{\bf S}_{jB}\rangle_{0}$ while the intra-sublattice spin exchange term gets renormalized with a different factor of $g_{s\alpha\alpha}$. The only other dimer term which does not get rescaled under the Gutzwiller projection is $H_d=  -\frac{t^2}{\Delta}\sum\limits_{<ij>,\sigma}[(1-n_{iA\bar{\sigma}})(1-n_{jB})+(n_{iA}-1)n_{jB\bar{\sigma}}]$, as it consists of only density operators~\cite{edegger-advphysics,Anwesha1}.
Then we have the important trimer terms: $H_{tr}=-\frac{t^2}{\Delta}\sum\limits_{<ijk>,\sigma}[g_{A\sigma}c_{kA\sigma}^{\dagger}n_{jB\bar{\sigma}}c_{iA\sigma}+g_{2}c_{iA\bar{\sigma}}c_{jB\bar{\sigma}}^{\dagger}c_{jB\sigma}c_{kA\sigma}^{\dagger}] -\frac{t^2}{\Delta}\sum\limits_{<jil>,\sigma}[g_{B\sigma}c_{lB\sigma}(1-n_{iA\bar{\sigma}})c_{jB\sigma}^{\dagger}+g_{2}c_{lB\sigma}c_{iA\sigma}^{\dagger}c_{iA\bar{\sigma}}c_{jB\bar{\sigma}}^{\dagger}] +\frac{tt'(U+\Delta)}{2U\Delta}\sum\limits_{<kj>,<<ik>>\sigma}\bigg[g_{t\sigma}c_{iA\sigma}^{\dagger}(1-n_{kA\bar{\sigma}})c_{jB\sigma}-g_{t\sigma}c_{jA\sigma}^{\dagger}n_{kB\bar{\sigma}}c_{iB\sigma}+g_{AAB\sigma}c_{iA\sigma}^{\dagger}c_{kA\bar{\sigma}}^{\dagger}c_{kA\sigma}c_{jB\bar{\sigma}}+g_{BBA\sigma}c_{jA\sigma}^{\dagger}c_{kB\bar{\sigma}}^{\dagger}c_{kB\sigma}c_{iB\bar{\sigma}}\bigg]+h.c.$

The various Gutzwiller factors involved (see \cite{Anwesha1} for details) are as follows: $g_{A\sigma}=2\delta/(1+\delta+\sigma m_{A})$,$g_{B\sigma}= 2\delta/(1+\delta-\sigma m_{B})$, $g_{t\sigma}=\sqrt{g_{A\sigma}g_{B\sigma}}$, $g_{s\alpha_{1}\alpha_{2}}=4/\sqrt{((1+\delta)^2-m_{\alpha_{1}}^2)((1+\delta)^2-m_{\alpha_{2}}^2)}$, $g_{2}=\delta g_{sAB}$,and $g_{\alpha_{1}\alpha_{1}\alpha_{2}\sigma} = 4\delta/\sqrt{((1+\delta)^{2}-m_{\alpha_{1}}^{2})(1+\delta+\sigma m_{\alpha_{1}})(1+\delta+\sigma m_{\alpha_{2}})}$ . Below we give details about the superconducting order parameter and the spectral functions.
 
  {\it Superconducting order parameter $\Phi_{d/s}$}: The SC correlation function is the two particle reduced density matrix defined by $F_{\gamma_1\gamma_2}({\bf{r}}_i-{\bf{r}}_j) =\la B^\dagger_{i\gamma_1}B_{j\gamma_2}\ra$ where $B^\dagger_{i\gamma} \equiv \Delta^{\gamma}_{AB}\equiv \langle c_{iA\uparrow}^{\dagger}c_{i+\gamma B\downarrow}^{\dagger}-c_{iA\downarrow}^{\dagger}c_{i+\gamma B\uparrow}^{\dagger}\rangle$ creates a singlet on the bond $(i,i+\gamma)$ where $\gamma$ is $x$ or $y$, considering d-wave pairing symmetry ($\Delta^x_{AB} = - \Delta^y_{AB} \equiv \Delta_d$) and extended s-wave pairing symmetry ($\Delta^x_{AB} =  \Delta^y_{AB} \equiv \Delta_s$) separately. The SC order parameter $\Phi_{d/s}$ is defined in terms of the  off-diagonal long-range order in this correlation $F_{\gamma_1,\gamma_2}({\bf{r}}_i-{\bf{r}}_j) \rightarrow \la B^\dagger_{i \gamma_1}\ra \la B_{j \gamma_2}\ra = \Phi_{\gamma_1} \Phi_{\gamma_2}$ as $|{\bf{r}}_i-{\bf{r}}_j| \rightarrow \infty$. Since $F_{\gamma_1\gamma_2}({\bf{r}}_i-{\bf{r}}_j)$ also corresponds to hopping of two electrons from $(j,j+\gamma_2)$ to sites $(i,i+\gamma_1)$, in the projected wavefunction scheme it scales just like the product of two hopping terms such that $F_{\gamma_1\gamma_2} \approx g_{A\uparrow}g_{B\downarrow}F_{\gamma_1\gamma_2}^0$. Hence the rescaled form of the superconducting order parameter is $\Phi_{d/s} \approx \sqrt{g_{A\uparrow}g_{B\downarrow}}\Phi_{d/s}^0$ where $\Phi_{d/s}^0 \equiv \Delta_{d/s}$ is the order parameter calculated in the unprojected wavefunction of the low energy effective Hamiltonian in Eq. (2). 
 
{\it Spectral Functions and Density of States:} In the main paper we also discussed the single particle density of states (DOS) and the spectral functions.
In the Gutzwiller projection method, the Green's function is rescaled with the appropriate Gutzwiller factor such that $G_{\alpha\sigma}(k,\omega)=g_{\alpha\sigma}G_{\alpha\sigma}^{0}(k,\omega)$ where $G_{\alpha\sigma}^{0}(k,\omega)$ is calculated in the unprojected basis. Here $\alpha$ represents the sublattice A or B and $\sigma$ is the spin index. The spectral function, $A_{\alpha\sigma}(k,\omega)$ which is imaginary part of the Green's function also get rescaled with the same Gutzwiller factors.
The results presented in the paper are for the spectral functions averaged over the two sublattices $A_\sigma(k,\omega)=\frac{1}{2}\sum_\alpha A_{\alpha\sigma}(k,\omega)$ which can be expressed as $A_\uparrow(k,\omega) = (|u_{1\uparrow k}|^2+|u_{2\uparrow k}|^2)\delta(\omega-E_{1\uparrow}(k))+(|u_{3\uparrow k}|^2+u_{4 \uparrow k}|^2)\delta(\omega-E_{2\uparrow}(k))+(|v_{1 \uparrow k}|^2+|v_{2\uparrow k}|^2)\delta(\omega+E_{1\downarrow}(k))+(|v_{3\uparrow k}|^2+|v_{4 \uparrow k}|^2)\delta(\omega+E_{2\downarrow} (k)) $. The down spin spectral function can be obtained by replacing $u_{i\uparrow k}\leftrightarrow v_{i\downarrow k}$ (and vice-versa) and by replacing $E_{i\sigma}(k)$ by $-E_{i\sigma}(k)$. 
Here $E_{1,2,\uparrow}(k)$ are the eigenvalues of the BdG equation for a given $k$ in the BZ with eigenvectors $(u_{1\uparrow k},u_{2\uparrow k},-v_{1\downarrow k},-v_{2\downarrow k})$ and $(u_{3 \uparrow k},u_{4 \uparrow k},-v_{3\downarrow k},-v_{4\downarrow k})$ respectively and $-E_{1,2\downarrow}$ are eigenvalues corresponding to eigenvectors obtained by $u_{i \sigma k} \rightarrow v_{i\sigma k}$ and $v_{i\sigma k} \rightarrow -u_{i\sigma k}$. In order to get the low energy spectral functions, presented in the main paper, we integrate $A_{\sigma}(k,\omega)$ over a small $\omega$ range such that $ |\omega| \le (0.01-0.02)t$.

The single particle density of states is defined as, $\rho_{\alpha\sigma}(\omega)=\sum_{k} A_{\alpha\sigma}(k,\omega)$.
The results presented in the paper are for the single particle density of states (DOS) in the up spin  and down spin channels, defined as $\rho_{\sigma}(\omega)=(\rho_{A\sigma}(\omega)+\rho_{B\sigma}(\omega))/2$.
The zero temperature momentum distribution function, which helps in identifying whether a Fermi pocket is an electron pocket or a hole pocket (and is presented in section SM F) can also be obtained from the spectral function using
$n_{\sigma}(k)= \int_{-\infty}^{0} d \omega A_{\sigma}(k,\omega)$.
\\
\\
  {\bf SM B. Details of the renormalized mean field theory}:
  \\
  \noindent
  We solve the renormalized effective low energy Hamiltonian using three different versions of the renormalized mean field theory (RMFT). To explore the SC phase, we use a generalised spin-symmetric Bogoliubov mean field theory, which basically maps onto a two-site Bogoliubov-deGennes (BdG) mean field theory for each allowed $k$ point in the BZ. We do a mean field decomposition of the various terms in the Hamiltonian, and self-consistently solve for the following mean fields :  (a) pairing amplitude, $\Delta^{\gamma}_{AB}\equiv \langle c_{iA\uparrow}^{\dagger}c_{i+\gamma B\downarrow}^{\dagger}-c_{iA\downarrow}^{\dagger}c_{i+\gamma B\uparrow}^{\dagger}\rangle$, where $\gamma$ is $x$ or $y$, considering d-wave pairing symmetry ($\Delta^x_{AB} = - \Delta^y_{AB} \equiv \Delta_d$) and extended s-wave pairing symmetry ($\Delta^x_{AB} =  \Delta^y_{AB} \equiv \Delta_s$) separately; (b) density difference between two sublattices, $\delta=(n_{A}-n_{B})/2$; (c) inter sublattice fock shifts, $\chi_{AB\sigma}^{(1)}=\langle c_{iA\sigma}^{\dagger}c_{jB\sigma}\rangle,j=i\pm x,i \pm y,\chi_{AB\sigma}^{(2)}=\langle c_{iA\sigma}^{\dagger}c_{jB\sigma}\rangle,j=i\pm 2x\pm y$ or $i \pm 2y \pm x$; and (d) intra sublattice fock shift on A(B) sublattice, with $\chi_{\alpha\alpha\sigma}=\langle c_{i\alpha\sigma}^{\dagger}c_{i\pm 2x/2y \alpha\sigma} $+h.c.$\rangle$, and $\chi_{\alpha\alpha\sigma}^{'}=\langle c_{i\alpha\sigma}^{\dagger}c_{i\pm x \pm y \alpha\sigma}$+h.c.$\rangle$. To explore the magnetic order and the phase transitions involved, we solve the renormalized Hamiltonian using standard mean field theory allowing non-zero values of the sublattice magnetization $m_{\alpha}=n_{\alpha\uparrow}-n_{\alpha\downarrow}$ with $\alpha=A,B$, from which one gets the staggered magnetization $m_s=(m_A-m_B)/2$ and  the uniform magnetisation $m_f=(m_A+m_B)/2$, along with all other mean-fields mentioned above except for the SC pairing amplitudes $\Delta_{s/d}$. The third calculation, where we allow for both the SC pairing amplitudes and the magnetization along with all other mean fields metioned above, uses a standard canonical transformation followed up by the Bogoliubov transformation to diagonalise the mean field Hamiltonian neglecting the inter-band pairing as weak. We solve the RMFT self-consistent equations on the square lattice for various values of $U,\Delta$ and $t^\prime$ to obtain the phase diagram reported in the paper. In the parameter regime where solutions with nonzero SC pairing amplitudes and magnetization (from the first two calculations) are both viable, we compare the ground state energy of the two mean-field solutions to determine the stabler ground state. We finally compare the energy of this state with the one obtained in the third calculation to determine the true ground state. Below we give details about the ground state energy comparisons.
 \renewcommand{\thefigure}{{\bf S1}}
\begin{figure}[h!]
  \begin{center}
    \includegraphics[scale=0.32,angle=-90]{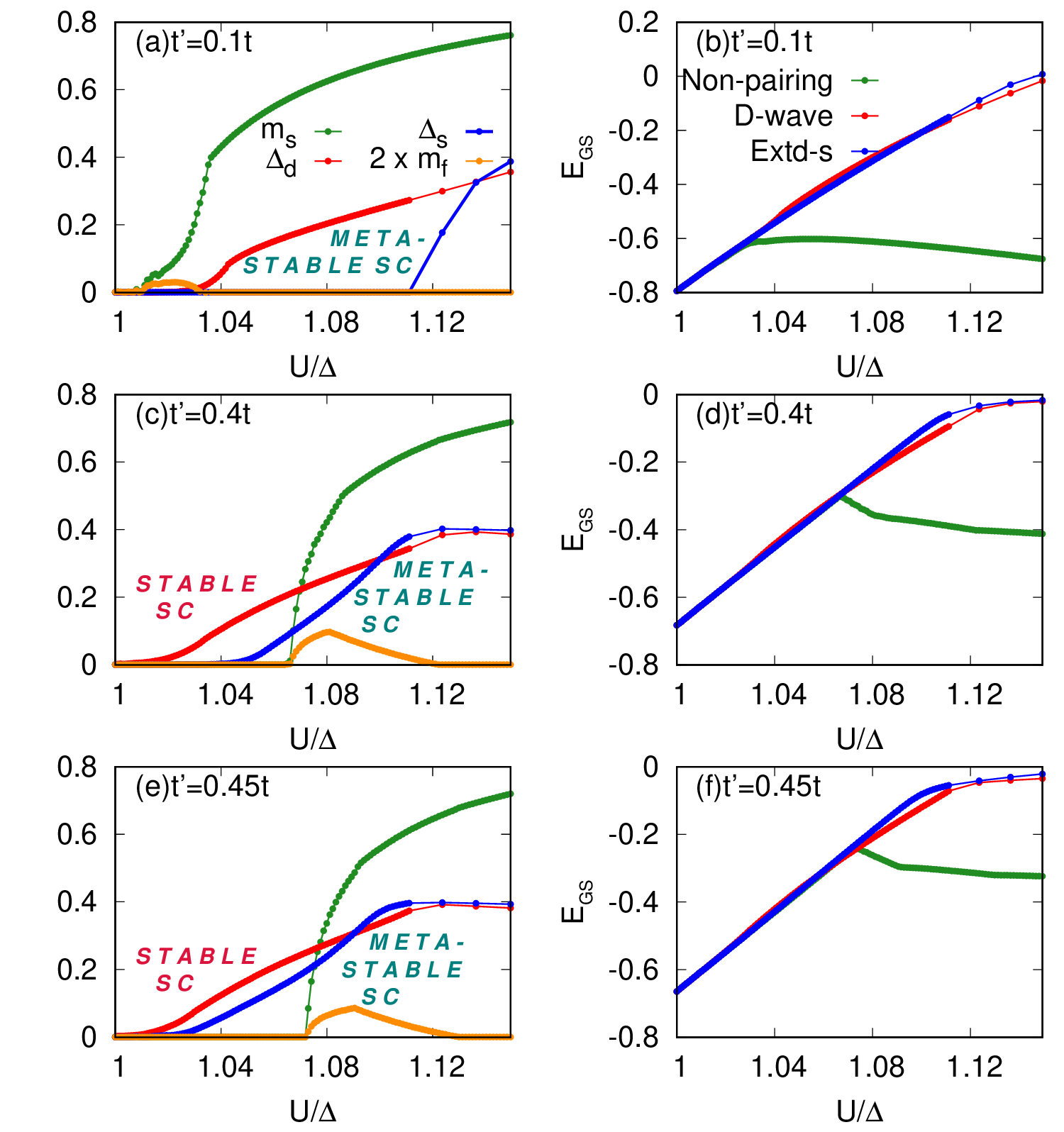}
    \vskip0.3cm
    \caption{Left panels show various mean fields, namely, the staggered magnetization $m_s$, uniform magnetization $m_f$, d-wave pairing amplitude $\Delta_d$ and the extended s-wave pairing amplitude $\Delta_s$ as functions of $U/\Delta$ for different values of $t^\prime$ at $U=10t$ for the 2d square lattice. Right panels show the ground state energies for the d-wave SC phase, extended s-wave SC phase and the non-superconducting phase where only magnetic order is allowed, as functions of $U/\Delta$.}
  \end{center}
\end{figure}
\\
{\bf SM C. Competing Order-Parameters and Ground State Energy Comparison}:
    \\
    \noindent
   Comparison of the results from the first two calculations shows that there is a significantly broad regime of parameters over which  the SC and magnetic orders both exist and compete with each other. In order to determine the true nature of the ground state in this parameter regime, we compare the ground state energies of the different RMFT solutions.
    As shown in Fig. S1, even for small values of $t^\prime$, the SC pairing amplitudes, in both the pairing channels studied, turn on but the magnetic transition precedes the transition into the SC phase. Once the magnetic order turns on, the ground state energy of the non-superconducting solution becomes lower than that of both the SC phases studied as shown in the right panels of Fig. S1. Thus for $t^\prime < 0.1t$ there is no stable SC phase, as shown in Fig (2e) of the main paper. For larger values of $t^\prime$, as $U/\Delta$ increases  superconductivity turns on before the magnetic order sets in. There continues to be a solution of the RMFT with pairing amplitudes, in either of the symmetry channels, non zero even in the magnetically ordered regime, but the non-superconducting magnetically ordered solution is lower in energy here. Thus the pure SC phase is a stable phase only before the magnetic transition point.
  \renewcommand{\thefigure}{{\bf S2}}
     \begin{figure}[h!]
  \begin{center}
    \includegraphics[scale=0.32,angle=-90]{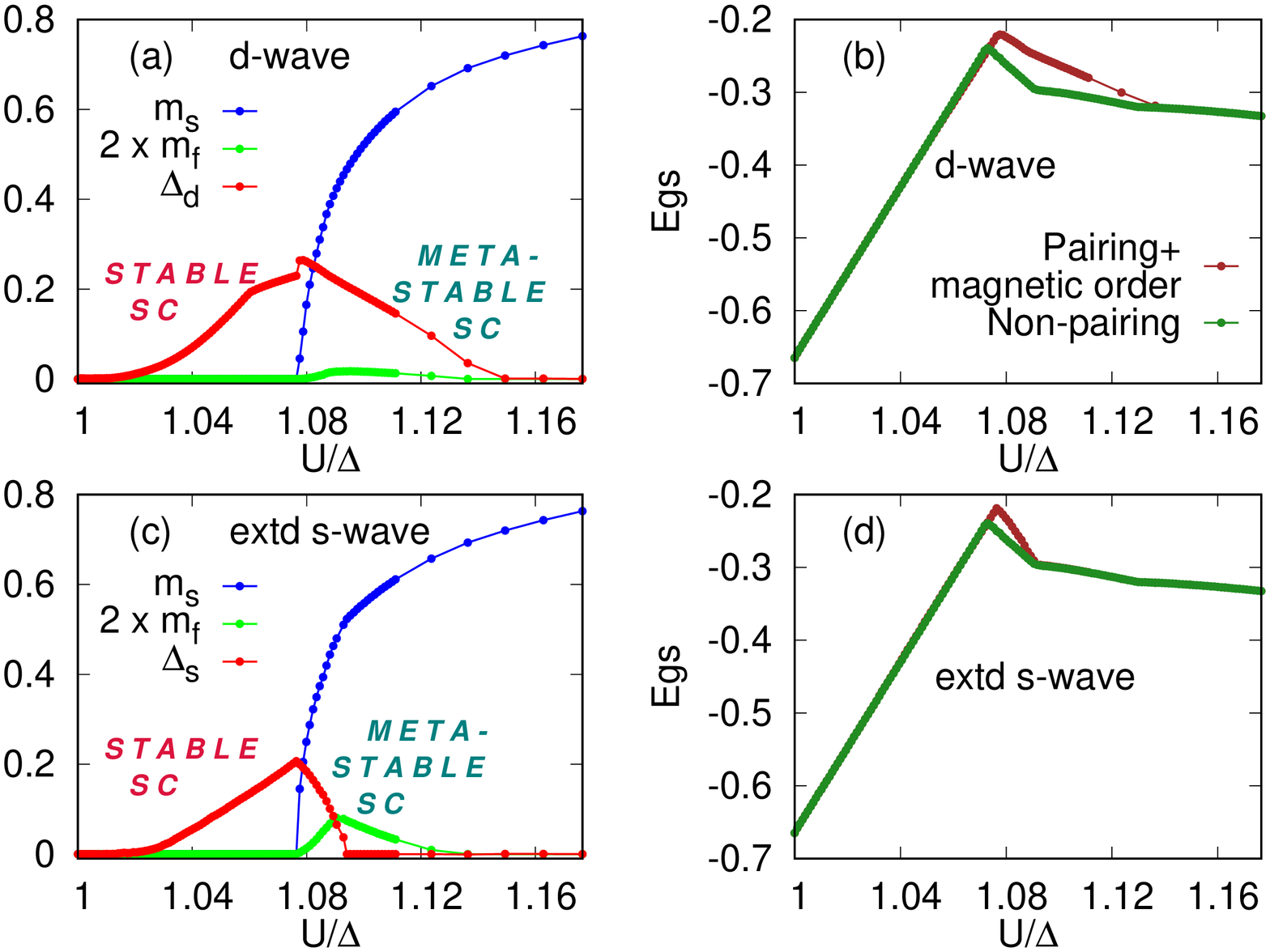}
    \caption{Top left panel shows several mean fields obtained from the third solution of the RMFT where both SC pairing and magnetic order are allowed, namely, the staggered magnetization $m_s$, uniform magnetization $m_f$, and the d-wave pairing amplitude $\Delta_d$ as functions of $U/\Delta$ for $t^\prime=0.45t$ and $U=10t$. Top right panel shows the ground state energy of the non-superconducting phase where only magnetic order is allowed and the energy for the third solution as functions of $U/\Delta$. Note that the phase with both orders coexisting is only a metastable phase. Lower panels show similar results for the extended s-wave SC order.}
  \end{center}
    \end{figure}

    There is a third scenario possible where one can do a RMFT allowing for non-zero values of both SC and magnetic order parameters along with other mean fields. Before the magnetic order turns on, this theory is consistent with the spin-symmetric Bogoliubov theory described above. After the magnetic order sets in, differences between the two calculations become visible. In the third calculation, the SC order coexists with the ferrimagnetic order for a range of parameters as shown in Fig. S2 though the pairing amplitudes decrease with increasing $U/\Delta$. Comparing the energy of this phase with that of the ferrimagnetic metal phase, which was found to be the stabler phase by comparing energies of first two calculations in this regime,  we find that the coexistence phase is also a metastable phase and the system actually stabilizes into the ferrimagnetic metallic phase as shown in Fig. 2, of the main paper.
    \\
    \\
    \noindent
        {\bf SM D.  Phase-diagram in $U/t-U/\Delta$ plane for a fixed $t^\prime$}:\\
        \noindent
        In the main paper we have shown phase-diagrams for the IHM on a 2d square lattice for a fixed value of $U/t$. Fig. 2[e] showed the phase diagram in $t^\prime/t-U/\Delta$ plane for a fixed $U$ and Fig. 1 showed a section of this phase diagram for $t^\prime=0.4t$.
        
\renewcommand{\thefigure}{{\bf S3}}
    \begin{figure}[h!]
  \begin{center}
  \includegraphics[scale=0.32,angle=-90]{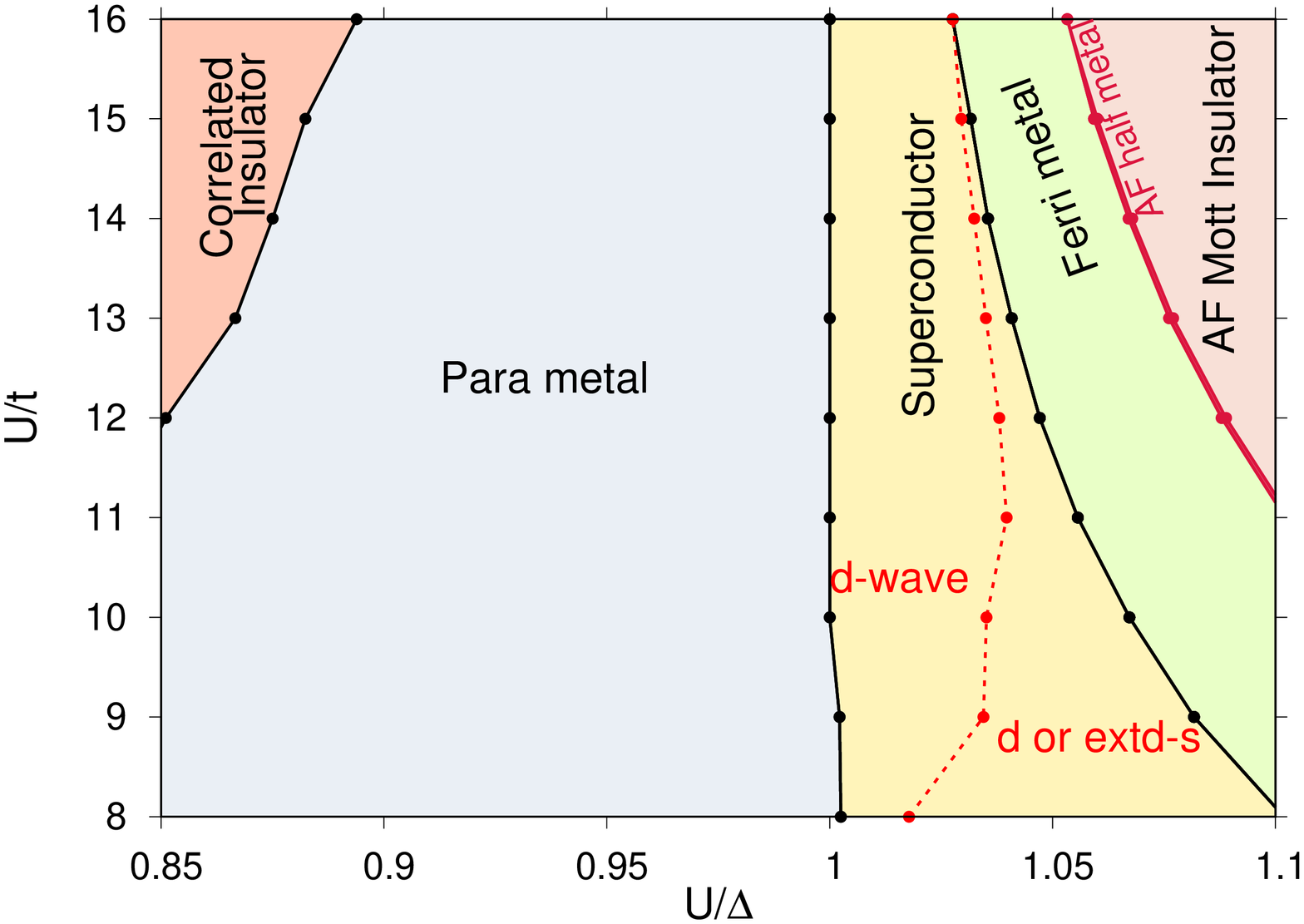}
  \caption{Phase diagram of the half-filled IHM on a 2d square lattice in $U/t-U/\Delta$ plane for $t^\prime=0.4t$. Note that the SC phase always turns on for $U \sim \Delta$ irrespective of the value of $U/t$ within the range of validity of the calculation. As $U/t$ increases, the range of $U/\Delta$ over which both the s-wave and the d-wave SC phases are viable solutions and almost degenerate shrinks rapidly while the range of $U/\Delta$ over which only the d-wave SC phase is stable reduces rather slowly.}
\end{center}
\end{figure}

        In order to understand how the different phases and the phase boundaries between them evolve with varying $U$, here we have shown the phase diagram in $U/t-U/\Delta$ plane for a fixed $t^\prime/t$. As shown in Fig. S3, superconductivity always turns on for $U\sim \Delta$ irrespective of the  value of $U/t$ though with increase in $U/t$, the range of $U/\Delta$ over which both pairing symmetries are almost degenerate solutions shrinks rapidly such that eventually, for large enough values of $U/t$, the system has only a d-wave SC phase.  
    \\
    \vskip0.25cm
      \noindent
      {\bf SM E.  Low Energy Spectral Functions with Varying $t^\prime$}:\\
      \noindent
        In order to understand the charge dynamics as the system approaches the SC phase with the tuning of second neighbour hopping, $t^\prime$, we have analysed the single particle spectral functions  for a fixed $U/\Delta$ in the ferrimagnetic metallic phase. Note that the main text showed how the low energy spectral functions $A_{\sigma} (k, \omega \sim 0)$ change with the tuning of $U/\Delta$ for a fixed $t^\prime$.
 \renewcommand{\thefigure}{{\bf S4}}
    \begin{figure}[h!]
  \begin{center}
  \includegraphics[scale=0.32]{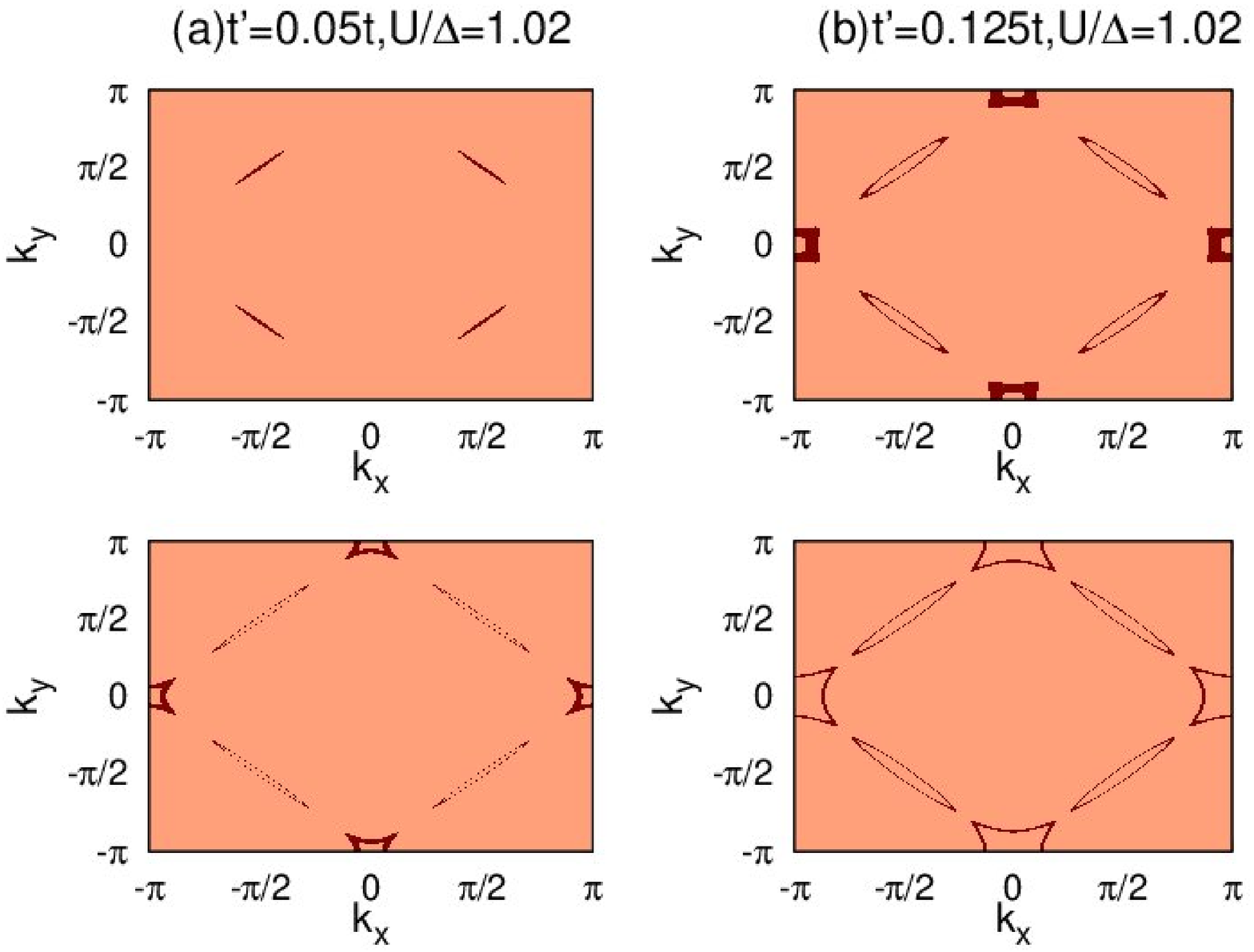}
  \caption{Here we show  the low energy spectral functions $A_\sigma(k,\omega \sim 0)$ (integrated over $|\omega|\le (0.01-0.02)t$ on a  $3000 \times 3000$ lattice) in the full Brillouin zone (BZ) for the ferrimagnetic phase at a fixed $U/\Delta=1.02$ and for two values of $t^\prime$. Upper panels show $A_{\uparrow}(k,\omega \sim 0)$, and the bottom panels $A_{\downarrow}(k,\omega \sim 0)$. }
\end{center}
    \end{figure}
   
   We can understand why the SC phase does not get stabilized for small values of $t^\prime$ by looking at the evolution of $A_{\sigma} (k, \omega \sim 0)$ for a fixed $U/\Delta$ as one tunes $t^\prime$. Fig. S4 shows $A_{\sigma} (k, \omega \sim 0)$ close to the magnetic transition point of $t^\prime=0$, that is, for $U/\Delta=1.02$. For small values of $t^\prime$, at this value of $U/\Delta$ the system is in the ferrimagnetic metal phase. As we increase $t^\prime$ inside the ferrimagnetic metal phase, the up spin spectral functions get bigger electron pockets around ${\bf{K}}=(\pm \pi/2,\pm \pi/2)$ points while the down spin spectral functions get bigger hole pockets around ${\bf{K}}^\prime=(\pm \pi,0),(0,\pm \pi)$ points.  In addition to this, as $t^\prime$ increases even the up-spin spectral functions get hole pockets and the down spin spectral functions get electron pockets. As a result of both these effects, an almost connected contour of Fermi pockets is formed, whence superconductivity emerges by the formation of Cooper pairs of the corresponding low energy quasiparticles.
    \\
    \vskip0.4cm
      \noindent
          {\bf SM F. Nature of Fermi Pockets in the Low Energy Spectral Functions and Momentum Distribution Functions}:\\
          \noindent
             \renewcommand{\thefigure}{{\bf S5}}
    \begin{figure}[h!]
      \begin{center}
    \includegraphics[scale=0.32]{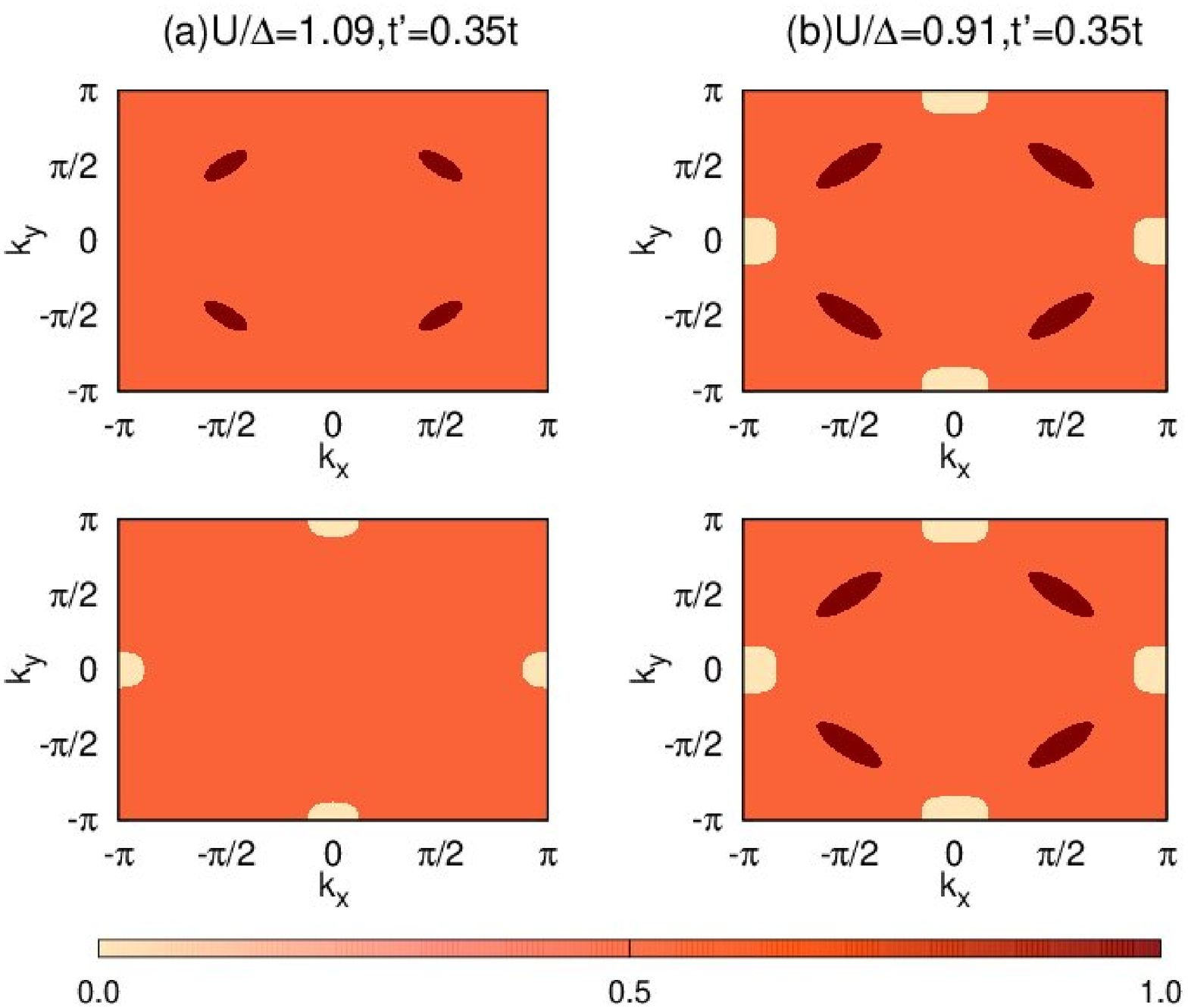}
    \caption{Momentum distribution function $n_\sigma(k)$ in the ferrimagnetic metal and the para metal phases for $t^\prime=0.35t$. In the ferrimagnetic metal phase shown in panel (a)  $n_\uparrow(k) > 1/2$ on (electron) pockets centered around the ${\bf{K}}$ points while $n_\downarrow(k) < 1/2$ on (hole) pockets centered around the ${\bf{K}}^\prime$ points in the BZ. Panel (b) shows the results for the paramagnetic metal phase, where the systen has spin symmetry and $n_\sigma(k) <1/2$ around the ${\bf{K}}^\prime$ points while $n_\sigma(k)>1/2$ around the ${\bf{K}}$ points for both the spin components. Everywhere else in the BZ $n_\sigma(k)=1/2$ in all the panels.}
    \vskip-1cm
\end{center}
\end{figure}

          The momentum distribution function $n_\sigma(k)$, defined in the section (SM A), is uniformly half in the entire BZ for any insulating phase of the model studied here. When the system goes into a metallic phase, at least one of the bands cross the Fermi level resulting in filled or empty Fermi pockets depending on the curvature of the band. Filled Fermi pockets, also called electron pockets, have $n_\sigma(k) > 1/2$, while empty Fermi pockets, also called hole pockets, have $n_\sigma(k) < 1/2$. Fig. S5 shows $n_\sigma(k)$ for $t^\prime=0.35t$ for two values of $U/\Delta$. Panel (a) shows the result for the ferrimagnetic metal phase and panel (b) shows the results in the para metal phase. In the ferri-metal phase, $n_{\uparrow}(k)$ has filled pockets around the ${\bf{K}}$ points while the down-spin component has hole pockets around the ${\bf{K}}^\prime$ points in the BZ. In the para-metal phase, shown in panel (b), there is a spin symmetry and $n_\sigma(k)$ has electron and hole pockets for both the spin channels. 
  \renewcommand{\thefigure}{{\bf S6}}
\begin{figure}[h!]
  \begin{center}
    \includegraphics[scale=0.23,angle=-90]{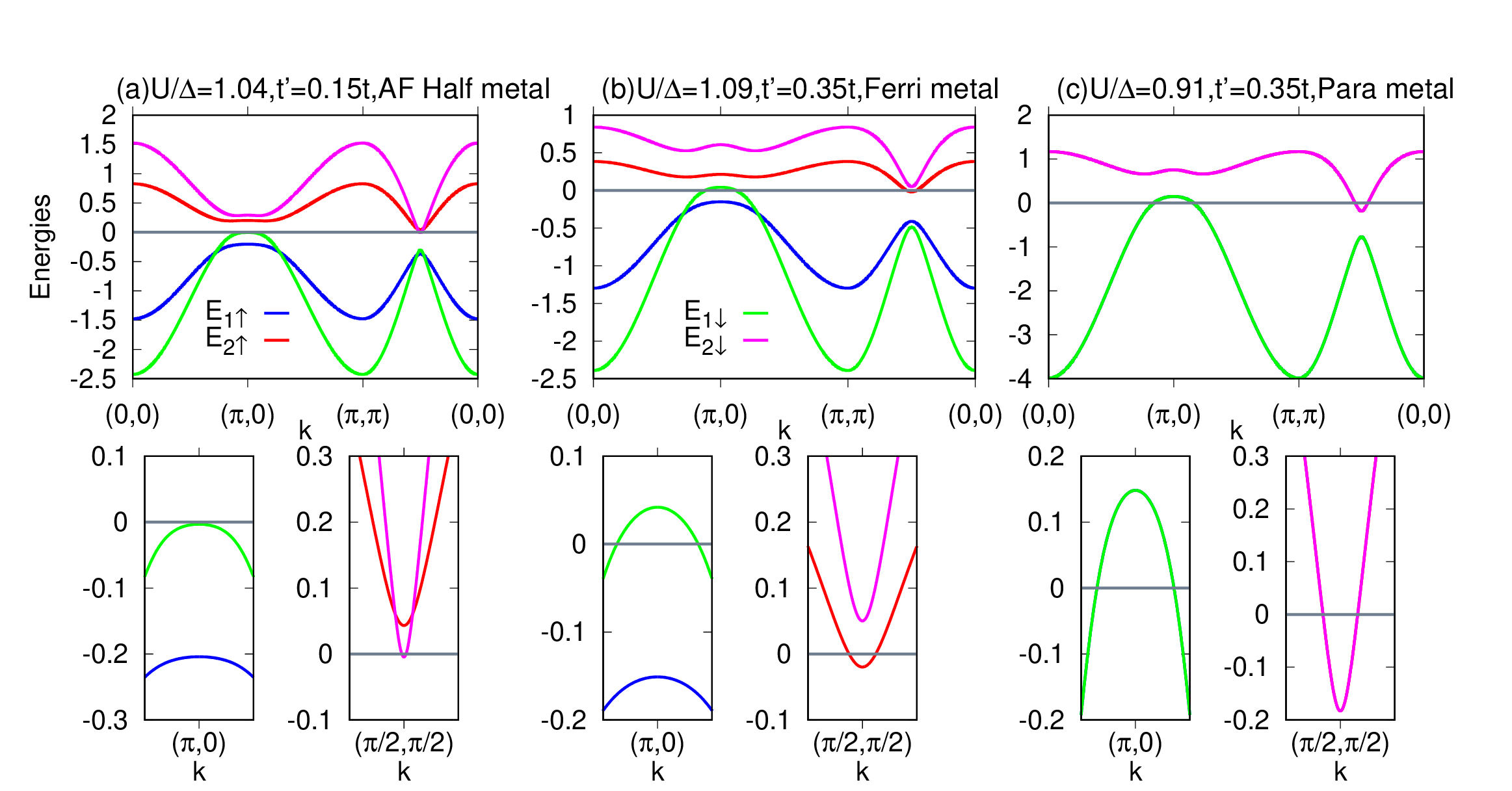}
    \caption{Band dispersion $E_{n\sigma}(k)$ on paths along high symmetry directions in the BZ. Panel (a) shows bands in the AF half-metal phase where both the down spin bands  cross the Fermi level near the ${\bf{K}}$ and ${\bf{K}}^\prime$ points while the up spin bands are fully gapped. Panel (b) shows bands in the ferrimagnetic metal phase, where one down-spin band crosses the Fermi level near the ${\bf{K}}^\prime$ points while one up-spin band crosses the Fermi level near the ${\bf{K}}$ point and the other two bands are gapped.  Panel (c) shows bands in the paramagnetic metal phase where there is a spin symmetry and all the bands cross the Fermi level. The lower panels zoom in close to the band crossing at the Fermi energy. }
\end{center}
\end{figure}
 Fig. S6 shows the band dispersion $E_{n\sigma}(k)$ for both the bands on paths along high symmetry directions in the BZ.  In the AF half-metal phase, the down spin channel has small hole pockets around ${\bf{K}}^\prime$ and tiny electron pockets around ${\bf{K}}$. In the Ferrimagnetic metal phase, the down spin band $E_{1\downarrow}(k)$ crosses the Fermi energy around the ${\bf{K}}^\prime$ points resulting in small hole pockets and $E_{2\uparrow}(k)$ crosses the Fermi energy near the ${\bf{K}}$ points resulting in small electron pockets. In the paramagnetic metal phase, $E_1(k)$ crosses the Fermi energy around the ${\bf{K}}^\prime$ points resulting in hole pockets and $E_2(k)$ crosses the Fermi level around ${\bf{K}}$ resulting in electron pockets, where, because of the spin symmetry, we have suppressed the spin indices.
\\
\vskip0.5cm
{\bf SM G. Spectral Functions in the AF Half-Metal Phase}:
\\
\renewcommand{\thefigure}{{\bf S7}}
\begin{figure}[h!]
  \begin{center}
    \includegraphics[scale=0.32,angle=-90]{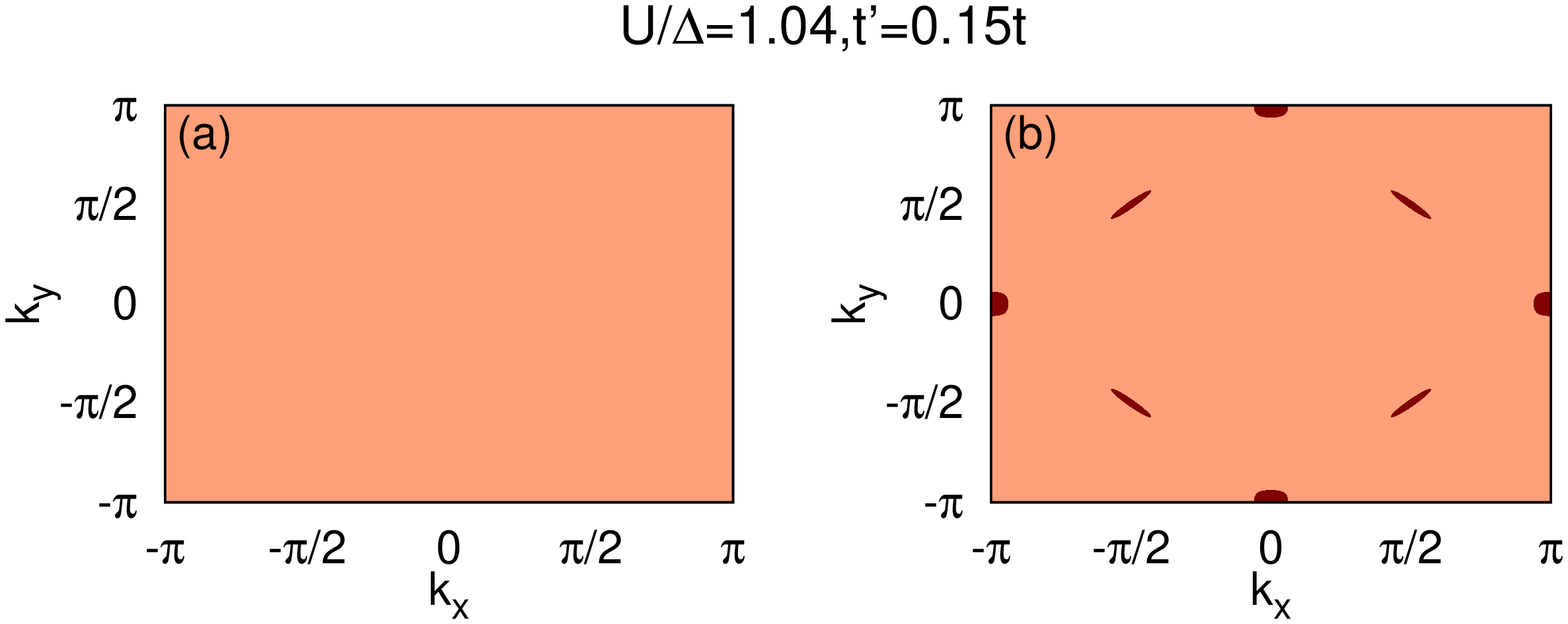}
  \caption{Spin resolved low energy spectral function $A_\sigma(k,\omega\sim 0)$ (integrated over $|\omega| \le 0.01t$) in the AF half-metal phase. Left (right) panel shows the spectral function for the up-spin (down-spin) channel.  }   
\end{center}
\end{figure}
Finally, we show the low energy spectral function $A_\sigma(k,\omega\sim 0)$ for the AF half-metal phase (see Fig. S7), which is fully consistent with the band-dispersions shown above. The up-spin channel is gapped while $A_\downarrow(k,\omega\sim 0)$ has tiny electron pockets at the ${\bf{K}}$ points and hole pockets at the ${\bf{K}}^\prime$ points in the BZ.


\begin{thebibliography}{999}
\bibitem{hubbard}
    Note that this is in contrast to the strongly correlated limit of the half-filled Hubbard model, where the kinetic energy gets fully projected out from the low energy Hilbert space, resulting in purely insulating phases.
\bibitem{PM}
  This PM phase is adiabatically connected to the metallic phase observed for weak to intermediate strength of $U/t$ as long as $U \sim \Delta$ and the system is constrained to be paramagnetic, as shown in earlier work on the IHM using DMFT and other approaches~\cite{garg_metal,metal,metal2,metal3}
\bibitem{coexist}
    Though there is a metastable state in which the SC phase coexists along with the ferrimagnetic order for a range of $U/\Delta$ after the magnetic transition (see SM for details), due to the really tiny Zeeman splitting ($\le 0.035t$ for $U=10t$) produced by the small uniform magnetization $m_f$ the possibility of a Fulde-Ferrel-Larkin-Ovchinnikov (FFLO) state seems unlikely~\cite{FFLO,FFLO2,FFLO3}.
\bibitem{note}
  Though we have studied the IHM on the simplest square lattice, a qualitatively similar phase diagram is expected on any bipartite lattice with changes involving appropriate symmetries, e.g., $d+id$ pairing symmetry on a honeycomb lattice.
\bibitem{note2}
  Though $t^\prime$ helps in the formation of the SC phase with pairing amplitudes living on the nearest neighbour bonds, there is no significant second neighbour pairing induced by $J^\prime$.
\end{thebibliography}
\end{document}